\documentclass[twocolumn,english,english,showpacs,preprintnumbers]{revtex4}
\usepackage[T1]{fontenc}
\usepackage[latin1]{inputenc}
\usepackage{babel}

\makeatletter

\providecommand{\LyX}{L\kern-.1667em\lower.25em\hbox{Y}\kern-.125emX\@}

\usepackage[T1]{fontenc}
\usepackage[latin1]{inputenc}
\usepackage{babel}

\makeatletter

\input{epsf}

\def\frac#1#2{{\textstyle{{#1}\over {#2}}}}

\def\lsim{\mathrel{\rlap{\lower4pt\hbox{\hskip1pt$\sim$}}
    \raise1pt\hbox{$<$}}}
\def\gsim{\mathrel{\rlap{\lower4pt\hbox{\hskip1pt$\sim$}}
    \raise1pt\hbox{$>$}}}
\def\sqr#1#2{{\vcenter{\vbox{\hrule height.#2pt
         \hbox{\vrule width.#2pt height#1pt \kern#1pt
         \vrule width.#2pt}
         \hrule height.#2pt}}}}

\def\beq{\begin{equation}}
\def\eeq{\end{equation}}
\def\beqa{\begin{eqnarray}}
\def\eeqa{\end{eqnarray}}

\makeatother

\makeatother
\begin{document}

\preprint{DF/IST-7.2002}

\title{Noncommutative Scalar Field Coupled to Gravity }

\author{O. Bertolami and L. Guisado}

\affiliation{Departamento de Física, Instituto Superior Técnico\\
 Av. Rovisco Pais 1, 1049-001 Lisboa, Portugal}

\date{\today{}}

\pacs{11.10.Lm, 04.50+h, 98.80-Cq}

\begin{abstract}
A model for a noncommutative scalar field coupled to gravity is proposed
via an extension of the Moyal product. It is shown that there are
solutions compatible with homogeneity and isotropy to first non-trivial
order in the perturbation of the star-product, with the gravity sector
described by a flat Robertson-Walker metric. We show that in the slow-roll
regime of a typical chaotic inflationary scenario, noncommutativity
has negligible impact. 
\end{abstract}
\maketitle

\section{introduction}

The idea of noncommuting spatial coordinates has been proposed long
ago by Snyder \cite{Snyder}, when Quantum Field Theory itself had
just been successfully established. More recently, it has been pointed
out by Seiberg and Witten \cite{SW}, that noncommutative geometry
arises naturally in certain limits of string theory, which has motivated
great interest in this topic. Given the Moyal deformation of the product
of functions, which defines a noncommutative algebra, field theories
have subsequently been constructed (see e.g. Refs. \cite{Szabo,DougNekra}
for reviews) even though issues like unitarity \cite{Bahns} and 
renormalizability
are not yet fully established. On the other hand, realistic particle
physics models have been considered, allowing to constrain, to some
extent, the value of the noncommutative parameter \cite{Hewett}.

These results rely mostly on a noncommutative algebra provided by
a new product of functions. Noncommutative geometry has been systematized
most remarkably by Connes \cite{Connes} and Woronowicz \cite{Woronowicz},
via the generalized concept of differential structure of generic (\( C^{\star } \))-algebras.
This formulation is believed to be the way to construct a noncommutative
version of gravity and even quantum gravity, via noncommutative differential
calculus \cite{Varilly}. In this work however, we shall not follow
this approach but rather derive our results from the noncommutative
algebra of a generalized multiplication law for tensors while leaving
the Hilbert-Einstein action unchanged. It is believed that this noncommutative
algebra approach may provide some insight into the quantum gravity
physics at the Planck scale.

Several authors have studied the impact of new physics in inflationary
models and, hence, on the gaussian character of energy density fluctuations
or on the isotropy of the observables. Noncommutativity of the coordinates
introduces a fundamentally new length scale whose imprint may be important
\cite{Lizzi1,Green}. As will be discussed, our approach is similar
to the study carried out in Ref. \cite{Lizzi2}, even though differences
in details lead to somewhat different conclusions. Most remarkably
we find that our perturbation approach allows for an homogeneous and
isotropic description of the impact of noncommutativity.

\section{Generalized Moyal Product}

Noncommutativity is usually introduced in Minkowski space via a noncommutative
Moyal product defined as \begin{eqnarray}
 &  & T*W\left( x\right) =\\
 &  & \sum _{n=0}^{\infty }\frac{\left( i/2\right) ^{n}}{n!}\theta ^{\alpha _{1}\beta _{1}}\ldots \theta ^{\alpha _{n}\beta _{n}}\left( T_{,\alpha _{1}\ldots \alpha _{n}}\right) \left( W_{,\beta _{1}\ldots \beta _{n}}\right) ,\nonumber 
\end{eqnarray}
 where \( T \) and \( W \) are generic tensors whose indices have
been suppressed, the primes denote partial derivatives and \( \theta ^{\alpha \beta } \)
is constant. Aiming to preserve Lorentz symmetry we regard \( \theta ^{\alpha \beta } \)
as an antisymmetric Lorentz tensor. This yields the commutator between
coordinates \begin{equation}
\label{commutator}
\left[ x^{\mu },x^{\nu }\right] =i\theta ^{\mu \nu }.
\end{equation}

In order to preserve general covariance we introduce instead the following
generalized Moyal product \begin{eqnarray}
 &  & T*W\left( x\right) =\label{Moyal_gen} \\
 &  & \sum ^{\infty }_{n=0}\frac{\left( i/2\right) ^{n}}{n!}\theta ^{\alpha _{1}\beta _{1}}\ldots \theta ^{\alpha _{n}\beta _{n}}\left( T_{;\alpha _{1}\ldots \alpha _{n}}\right) \left( W_{;\beta _{1}\ldots \beta _{n}}\right) ,\nonumber 
\end{eqnarray}
 where the semicolon denotes covariant derivative with the Levi-Civitta
connection and \( \theta ^{\alpha \beta } \) is a non-constant rank-2
antisymmetric tensor. Despite being non-associative in general, we
shall see that for a scalar field \( \Phi  \) such that \( \theta ^{\alpha \beta }\Phi _{;\alpha }=0 \)
we recover associativity of the product to some extent.

By making use of the antisymmetry of \( \theta ^{\alpha \beta } \)
one can easily prove that, under conjugation, \( \left( T*W\right) ^{*}=W^{*}*T^{*} \).
The compatibility of the metric yields \( g^{\mu \nu }*T=g^{\mu \nu }T \)
so that the process of raising and lowering indices is not affected
by noncommutativity.

Noncommutative Lagrangian densities are obtained by turning usual
products into star-products so that one has to evaluate integrals
of the form \begin{equation}
S=\int d^{4}x\sqrt{-g}T^{*}*W.
\end{equation}

Through the process of integration by parts and dropping surface terms
we can arrange the covariant derivatives on the star-product to act
either on \( T \) or on \( W \), that is \begin{eqnarray}
S & = & \int d^{4}x\sqrt{-g}T^{*}\left( {\mathcal{A}}W\right) \\
 & = & \int d^{4}x\sqrt{-g}\left( {\mathcal{A}}T\right) ^{*}W,\nonumber 
\end{eqnarray}
 where \( {\mathcal{A}} \) is an Hermitian operator given by

\begin{eqnarray}
 &  & {\mathcal{A}}W=\\
 &  & \sum _{n=0}^{\infty }\frac{\left( -i/2\right) ^{n}}{n!}\left[ \theta ^{\alpha _{1}\beta _{1}}\ldots \theta ^{\alpha _{n}\beta _{n}}\left( W_{;\beta _{1}\ldots \beta _{n}}\right) \right] _{;\alpha _{n}\ldots \alpha _{1}}.\nonumber 
\end{eqnarray}

If the Lagrangian density is quadratic on the tensor \( T \) one
can use the property under conjugation to prove that \( T*T \) is
real; hence

\begin{eqnarray}
S' & = & \int d^{4}x\sqrt{-g}T*T=\int d^{4}x\sqrt{-g}T\left( {{\mathcal{A}}+{\mathcal{A}}^{*}\over 2}\right) T\nonumber \\
 & = & \int d^{4}x\sqrt{-g}T{\mathcal{O}}T,
\end{eqnarray}
 where the Hermitian operator \( {\mathcal{O}}=\frac{1}{2}\left( {\mathcal{A}}+{\mathcal{A}}^{*}\right)  \)
has been introduced

\begin{eqnarray}
 &  & {\mathcal{O}}W=\label{op_o} \\
 &  & \sum _{n=0}^{\infty }\frac{\left( -1/4\right) ^{n}}{\left( 2n\right) !}\left[ \theta ^{\alpha _{1}\beta _{1}}\ldots \theta ^{\alpha _{2n}\beta _{2n}}\left( W_{;\beta _{1}\ldots \beta _{2n}}\right) \right] _{;\alpha _{2n}\ldots \alpha _{1}}.\nonumber 
\end{eqnarray}

\section{noncommutative Scalar Field coupled to gravity}

\subsection{Massive scalar field}

The noncommutative action for a massive scalar field, \( \Phi  \),
is quadratic, and so, according with the results from the last section
\begin{equation}
S=-\frac{1}{2}\int d^{4}x\sqrt{-g}\left\{ \nabla ^{\mu }\Phi {\mathcal{O}}\nabla _{\mu }\Phi -m^{2}\Phi {\mathcal{O}}\Phi \right\} ,
\end{equation}
 and the equation of motion being given by \begin{equation}
\label{motion_fi}
\nabla ^{\mu }{\mathcal{O}}\nabla _{\mu }\Phi +m^{2}{\mathcal{O}}\Phi =0.
\end{equation}

Operator \( {\mathcal{O}} \) naturally arises in the equation of
motion. Furthermore, being an Hermitian operator it must correspond
to an observable of the scalar field \( \Phi  \). In the commutative
limit, \( lim_{\theta \rightarrow 0}{\mathcal{O}}=1 \). On the other
hand, switching off gravity and admitting that \( \theta ^{\alpha \beta } \)
is constant yields \( {\mathcal{O}}=1 \), since the partial derivatives
commute and are contracted with the antisymmetric tensor \( \theta ^{\alpha \beta } \)
in (\ref{op_o}). Hence, in this model, noncommutativity arises only
for non-trivial gravity. This is due to the fact that the usual Moyal
product obeys, under integration, the cyclic property

\begin{equation}
\int d^{4}x\, f*g=\int d^{4}x\, f\, g=\int d^{4}x\, g*f
\end{equation}

\noindent and the action we are considering is quadratic.

\subsection{Scalar field with an arbitrary potential}

In this section we consider the noncommutative generalization of an
arbitrary analytic commutative potential \( V\left( \Phi \right)  \).
Associativity played no part in the previous section because one dealt
with a quadratic action. Now, however, the case of an arbitrary potential
requires special care as, in general, the resulting star-product is
not associative. 

Given a commutative analytic potential \begin{equation}
V\left( \Phi \right) =\sum ^{\infty }_{n=0}{\lambda _{n}\over n!}\Phi ^{n}
\end{equation}
 one wishes to consider a noncommutative potential \( V_{NC}\left( \Phi \right)  \)
in the form

\begin{equation}
V_{NC}\left( \Phi \right) =\sum _{n=0}^{\infty }{\lambda _{n}\over n!}\overbrace{\Phi *\ldots *\Phi }^{n\: factors}.
\end{equation}
provided that the action of the star-product upon powers of the scalar
field \( \Phi  \) is associative. We shall analyze this generalization
of the potential in the case where \( \theta ^{\alpha \beta }\Phi _{;\beta }=0 \).

Since there is no a priori associativity, let us consider the sequence\begin{equation}
s_{2}=\left( \Phi *\Phi \right) \qquad s_{n+1}=\Phi *s_{n},\; \; \; n>2.
\end{equation}

It is easy to prove that, up to second order\begin{equation}
s_{n}\simeq \Phi ^{n}+{n\left( n-1\right) \over 2}\Phi ^{n-2}\left( \Phi \hat{*}\Phi \right) 
\end{equation}
where one defines\begin{equation}
\varphi \hat{*}\chi =-\frac{1}{8}\theta ^{\alpha _{1}\beta _{1}}\theta ^{\alpha _{2}\beta _{2}}\left( \varphi _{;\alpha _{1}\alpha _{2}}\right) \left( \chi _{;\beta _{1}\beta _{2}}\right) .
\end{equation}

Also, for every \( m \) and \( n \), it can be shown that, up to
second order\begin{equation}
s_{n}*s_{m}\simeq s_{m+n}
\end{equation}
which proves that one can calculate the power \( s_{q} \) grouping
\( q \) star-products in any way one wishes. Hence the star-product
of powers of \( \Phi  \) is associative under these conditions. Also,
these results enables one to write

\begin{equation}
\label{ncpot}
V_{NC}\left( \Phi \right) \equiv V\left( \Phi \right) +\frac{1}{2}V''\left( \Phi \right) \left( \Phi \hat{*}\Phi \right) ,
\end{equation}
 where \( '=d/d\Phi  \). 

The variation of the potential \( V_{NC} \) in the action is given
by

\begin{equation}
-{\delta S_{pot}\over \delta \Phi }=V'\left( \Phi \right) +\frac{1}{2}V'''\left( \Phi \right) \left( \Phi \hat{*}\Phi \right) -\frac{1}{4}{\mathcal{F}}\left[ V,\Phi \right] ,
\end{equation}
 where we have defined the operator

\begin{equation}
{\mathcal{F}}\left[ V,\Phi \right] =\left[ \frac{1}{2}V''\theta ^{\alpha _{1}\beta _{1}}\theta ^{\alpha _{2}\beta _{2}}\phi _{;\beta _{1}\beta _{2}}\right] _{;\alpha _{2}\alpha _{1}}.
\end{equation}

With this definition we also obtain that \begin{equation}
{\mathcal{O}}\Phi _{;\mu }\simeq \Phi _{;\mu }-\frac{1}{8}{\mathcal{F}}\left[ \Phi ^{2},\Phi _{;\mu }\right] .
\end{equation}

\subsection{Homogeneous and isotropic space-time}

In what follows we shall assume that the Einstein-Hilbert action is
unchanged by noncommutativity. This is done as our aim is the study
of the implications of a noncommutative algebra of tensors and because
there is no canonical way of introducing this algebra in the geometrical
formulation of gravity (i.e. in the Riemann tensor and, ultimately,
in the Ricci scalar). Hence the Einstein equations are given by \begin{equation}
\label{motion_grav}
R_{\alpha \beta }=-8\pi k\left[ \frac{1}{2}\nabla _{\{\alpha }\Phi {\mathcal{O}}\nabla _{\beta \}}\Phi +g_{\alpha \beta }V_{NC}\left( \Phi \right) \right] .
\end{equation}

As a concrete model we consider a homogeneous and isotropic space-time
described by the spatially flat Robertson-Walker metric \begin{equation}
ds^{2}=-dt^{2}+R^{2}\left( t\right) \left( dx^{2}+dy^{2}+dz^{2}\right) .
\end{equation}
 The non-vanishing components of the Christoffel symbols are given
by \begin{equation}
\Gamma _{\: ij}^{t}=R\dot{R}\delta _{ij}\qquad \Gamma _{\: jt}^{i}={\dot{R}\over R}\delta _{\: j}^{i}
\end{equation}
 and the Ricci tensor is diagonal and given by \begin{equation}
R_{tt}=3{\ddot{R}\over R}\qquad R_{ij}=-\left( R\ddot{R}+2\dot{R}^{2}\right) \delta _{ij}.
\end{equation}

The non-vanishing components of the antisymmetric noncommutative tensor,
\( \theta ^{\alpha \beta } \), correspond to two 3-vectors which
we denote by \( \vec{E} \) and \( \vec{B} \), in analogy with the
electromagnetic tensor. Thus, even if \( \theta ^{\alpha \beta } \)
is homogeneous, \( \theta ^{\alpha \beta }=\theta ^{\alpha \beta }\left( t\right)  \),
it is still possible that symmetry under rotations is broken and some
care should be taken concerning the choice of an isotropic \textit{Ansatz}
\textit{\emph{of the metric}}\textit{,} \textit{\emph{since \( \vec{E} \)
and \( \vec{B} \) can induce preferred directions in space. We shall
show, however, that there is a noncommutative model consistent with
homogeneity and isotropy to first order in perturbation theory, for
the homogeneous scalar field \( \partial _{i}\Phi =0 \). Under these
conditions, we get from Eqs. (\ref{motion_fi}) and (\ref{motion_grav})}}

\textit{\begin{eqnarray}
 &  & \ddot{\Phi }+3{\dot{R}\over R}\dot{\Phi }+V'=\label{mov_pert_1} \\
 &  & {\partial _{t}\left( R^{3}{\mathcal{F}}\left[ \Phi ^{2},\Phi _{;t}\right] \right) \over 8R^{3}}+\frac{1}{2}V'''\left( \Phi \hat{*}\Phi \right) +\frac{1}{4}{\mathcal{F}}\left[ V,\Phi \right] ,\nonumber \\
 &  & \nonumber \\
 &  & \nonumber \\
 &  & \left( {\dot{R}\over R}\right) ^{2}=\label{mov_pert_2} \\
 &  & {8\pi k\over 3}\left( \frac{1}{2}\dot{\Phi }^{2}+V+\frac{1}{2}V''\left( \Phi \hat{*}\Phi \right) -\frac{1}{16}\dot{\Phi }{\mathcal{F}}\left[ \Phi ^{2},\dot{\Phi }\right] \right) .\nonumber 
\end{eqnarray}
}

\textit{\emph{We show in the Appendix the explicit computation of
one of these terms, the remaining ones being analogous \begin{equation}
\label{nc_terms}
\begin{array}{c}
\Phi \hat{*}\Phi =-{1\over 2}\left( R\dot{R}\dot{\Phi }B\right) ^{2},\\
{\mathcal{F}}\left[ V,\Phi \right] ={1\over 2R^{3}}\partial _{t}\left[ R^{5}\dot{R}^{2}\dot{\Phi }B^{2}\frac{1}{2}V''\right] ,\\
{\mathcal{F}}\left[ \Phi ^{2},\dot{\Phi }\right] =-{2\over R^{3}}\partial _{t}\left[ R^{6}\dot{R}^{2}B^{2}\partial _{t}\left( {\dot{\Phi }\over R}\right) \right] ,
\end{array}
\end{equation}
 where we used the condition \( \vec{E}=0 \) \cite{Unit}. This condition
ensures that we have \( \theta ^{\alpha \beta }\Phi _{;\beta }=0 \)
and the noncommutative generalization of the scalar potential (\ref{ncpot})
makes sense. Hence we see that the dependence of Eqs. (\ref{nc_terms})
in \( \theta ^{\alpha \beta } \) is only through \( B^{2} \) and
consequently invariance under rotations is preserved. Since there
is no known dynamics for the \( \vec{B} \) field, we consider the
relationship}}

\textit{\emph{\begin{equation}
B^{2}=\hat{B}^{2}R^{-2\varepsilon },
\end{equation}
 where \( \hat{B}^{2} \) is a constant. We shall determine the parameter
\( \varepsilon  \) in the next section.}}

\section{\textit{\emph{Slow-roll in Chaotic Inflation}}}

\textit{\emph{Here we investigate the implications of noncommutativity
in the slow-roll phase of a typical chaotic inflation. Given its generic
features and the fairly general conditions for the onset of inflation,
chaotic models \cite{Linde} are particularly suited for studying
the effect of noncommutativity. We seek solutions of Eqs. (\ref{mov_pert_1})
and (\ref{mov_pert_2}) in first order of perturbation theory in \( \hat{B}^{2} \).
To perform this we consider solutions of the following form \begin{equation}
\Phi =\phi +\hat{B}^{2}\varphi \qquad R=a+\hat{B}^{2}\chi ,
\end{equation}
 where \( \phi  \) and \( a \) are solutions of the unperturbed
(commutative) problem, while \( \varphi  \) and \( \chi  \) are
arbitrary time dependent functions to be determined. We ignore in
every step higher order terms in \( \hat{B}^{2} \). Using units in
which \( k=1 \), Eqs. (\ref{mov_pert_1}) and (\ref{mov_pert_2})
take compact form \begin{eqnarray}
 &  & \ddot{\Phi }+3{\dot{R}\over R}\dot{\Phi }+V'=\hat{B}^{2}f,\\
 &  & \left( {\dot{R}\over R}\right) ^{2}={8\pi \over 3}\left( \frac{1}{2}\dot{\Phi }^{2}+V\right) +{8\pi \over 3}\hat{B}^{2}g\label{eqmotion} 
\end{eqnarray}
 in terms of functions \( f \) and \( g \) specified below. Standard
perturbation theory gives rise to the usual inflationary equations
\begin{eqnarray}
 &  & \ddot{\phi }+3{\dot{a}\over a}\dot{\phi }+V'\left( \phi \right) =0,\label{slow_eqs} \\
 &  & \left( {\dot{a}\over a}\right) ^{2}={8\pi \over 3}\left( {1\over 2}\dot{\phi }^{2}+V\left( \phi \right) \right) .\label{Fried} 
\end{eqnarray}
}}

\textit{\emph{Onset of inflation and slow-roll regime are achieved
once the following conditions are met}}

\textit{\emph{\begin{equation}
\label{slow_cond}
{V'\over V}\leq \sqrt{48\pi }\quad ,\quad {V''\over V}\leq 24\pi ,
\end{equation}
 so that we can neglect \( \ddot{\phi } \) in the Eq. (\ref{slow_eqs})
and \( \dot{\phi }^{2}/2 \) in Eq. (\ref{Fried}). It then follows
the useful condition \begin{equation}
\label{maj_fipto}
\left| \dot{\phi }\right| \leq \sqrt{2}V^{1/2}.
\end{equation}
}}

\textit{\emph{Therefore, terms in Eqs. (\ref{nc_terms}) can be evaluated
using the slow-roll conditions and, as illustrated in the Appendix,
we obtain that all of them are proportional to \( a^{4-2\varepsilon } \)
and to factors that depend on \( V \) and \( \dot{\phi } \) . Since
the Universe is expanding at an exponential rate, the perturbation
theory is meaningful only if \( \varepsilon \geq 2 \). However, \( \varepsilon >2 \)
implies that the terms in Eqs. (\ref{nc_terms}) decrease so rapidly
that noncommutativity will not lead to any effect. Thus, we conclude
from the consistency of perturbation theory that \( \varepsilon =2 \).
This is a natural choice from the theoretical point of view as well.
Most of the studied noncommutative models use a constant \( \theta ^{\alpha \beta } \).
If one requires that this is so for the physical coordinates \( y^{i}=R\, x^{i} \),
then one finds, inspired in Eq. (\ref{commutator}), \( \left[ y^{i},y^{j}\right] =\hat{B}^{ij} \)
and hence \( \varepsilon =2 \).}}

\textit{\emph{Equations for the perturbations are obtained gathering
all terms proportional to \( \hat{B}^{2} \) and this constant cancels
out from the differential equations. Function \( \varphi  \) satisfies
the relationship \begin{eqnarray}
 &  & f-4\pi {\dot{\phi }\over \dot{a}/a}g=\label{eqvarphi} \\
 &  & \ddot{\varphi }+3{\dot{a}\over a}\left[ 1+{4\pi \over 3}\left( {\dot{\phi }\over \dot{a}/a}\right) ^{2}\right] \dot{\varphi }+V''\left[ 1+4\pi {V'\over V''}{\dot{\phi }\over \dot{a}/a}\right] \varphi ,\nonumber 
\end{eqnarray}
 where functions \( f \) and \( g \) can be estimated (see Appendix)
using the slow-roll conditions: \begin{eqnarray}
 &  & \left| f\right| \leq a_{1}\frac{1}{2}V^{2}V''+a_{2}\frac{1}{2}V^{2}V'''+a_{3}V^{3}+a_{4}V^{5/2},\nonumber \\
 &  & \left| g\right| \leq a_{5}V^{3}+a_{6}\frac{1}{2}V^{2}V'',
\end{eqnarray}
 with \( a_{1}\simeq 85.5 \) , \( a_{2}\simeq a_{6}\simeq 4.2 \)
, \( a_{3}\simeq 3.30\times 10^{3} \) , \( a_{4}\simeq 4.52\times 10^{3} \)
and \( a_{5}\simeq 1.76\times 10^{2} \).}}

\textit{\emph{Potentials in chaotic inflation are characterized by
a small overall coupling constant, \( \lambda \simeq 10^{-14} \),
to ensure consistency with the amplitude of energy density perturbations
around \( 10^{-5} \), for \( \phi  \) field values of a few Planck
units. Writing the potential as \begin{equation}
\label{pot}
V\left( \Phi \right) =\lambda \, v\left( \Phi \right) ,
\end{equation}
 and considering \begin{equation}
\label{maj_pot}
v\leq 10^{2},
\end{equation}
 it then follows that \( \left| f\right| \leq 4.5\times 10^{-27} \)
and \( \left| g\right| \leq 1.8\times 10^{-34} \). However, the second
term of the right-hand side of Eq. (\ref{eqvarphi}) is of the order
\( 2.5\times 10^{-6} \) while the third term is of the order \( 7\times 10^{-11} \).
So, for numerical purposes, the left-hand side of the Eq. (\ref{eqvarphi})
is vanishingly small. But, in this case, one gets essentially the
same differential equation that would arise when performing perturbation
theory on the standard slow-roll approximation with no extra physics.
Therefore, we conclude that noncommutativity introduces no change
in inflationary slow-roll physics for the inflaton field.}}

\textit{\emph{Furthermore, from the equation for the \( \chi  \)
perturbation \begin{equation}
\label{eqchi}
{d\over dt}\left( {\chi \over a}\right) ={4\pi \over 3\dot{a}/a}\left( \dot{\phi }\dot{\varphi }+V'\varphi +g\right) 
\end{equation}
 we can see that the upper limit for \( \left| g\right|  \) assures
that, numerically, there is no difference in this equation as well.
Hence, we are led to conclude that performing perturbation calculations
on the presence of noncommutativity has no effect in the standard
chaotic inflationary model.}}

\section{\textit{\emph{Conclusions}}}

\textit{\emph{In this work we have considered a natural extension
of the Moyal product and studied its implications to the physics of
a scalar field coupled to gravity. We presented the general features
of our formalism and applied it to a specific study of the scalar
field on a spatially flat Robertson-Walker metric.}}

\textit{\emph{Our results were obtained using perturbation theory,
which necessarily requires that the antisymmetric noncommutative tensor,
\( \theta ^{\alpha \beta } \), is small compared to the covariant
derivative of the fields. This can be seen analyzing the explicit
expression for the Moyal product, Eq. (\ref{Moyal_gen}). Despite
the fact that there is no equation for \( \theta ^{\alpha \beta } \),
both perturbation theory and theoretical considerations show that
\( \theta ^{\alpha \beta }\sim R^{-2} \), where \( R \) is the scale
factor. So we find that once perturbation theory is valid, it remains
so as far as the expansion of the Universe lasts.}}

\textit{\emph{The antisymmetric tensor \( \theta ^{\alpha \beta } \)
can be parameterized by two three-vectors, just like in the case of
the electromagnetic tensor (c.f. Eq. (\ref{theta_param}) below).
The homogeneity requirement, that is, \( \partial _{i}\theta ^{\alpha \beta }=0 \),
could still lead to preferred directions in space rendering the isotropic
Ansatz of the Robertson-Walker metric meaningless. We show, in first
order perturbation theory, this is not the case since the terms arising
from noncommutative contributions depend only on the rotationally
invariants \( E^{2} \) and \( B^{2} \).}}

\textit{\emph{In the slow-roll regime in the context of a typical
chaotic inflation we show that noncommutativity introduces negligible
corrections. This is mainly due to two reasons. First, the scale parameter
(that might yield large deviations) does not appear in the first order
terms as \( \theta ^{\alpha \beta }\sim R^{-2} \), otherwise these
would grow exponentially and perturbation theory would be meaningless.
On the other hand, the slow-roll conditions yield small derivatives
for the inflaton field, Eq. (\ref{maj_fipto}), and for the logarithm
of the scale factor, Eq. (\ref{Fried}), given the small coupling
constant of the inflaton potential. Since the Moyal product involves
many derivatives (it is highly non-local) the smallness of the noncommutative
contributions is well understood. We can look at this from another
perspective: since perturbation theory requires that \( \theta ^{\alpha \beta } \)
is small compared to the derivatives and these are very small,
then noncommutativity is naturally suppressed. Therefore we are led
to conclude that noncommutative effects, if any, necessarily arise
in the non-perturbative regime of the theory.}}

Some remarks on our perturbative approach are in order. Our calculations
assume that perturbation theory is valid from a given cosmological
time \( t_{*} \) onward. Thus, if the conditions for inflation are
met and assuming that \textit{\emph{\( B=\hat{B}R^{-2} \),}} noncommutativity
plays a negligible role. This implies that \( B_{*}=\hat{B}R_{*}^{-2}\ll 1 \),
that is \( \hat{B}\ll R^{2}_{*} \), and therefore, a sufficiently
small \( \hat{B} \) garantees the validity of perturbation theory
for any given \( R_{*} \). The actual value for \( \hat{B} \) can
be quite small if \( t_{*} \) is the onset of inflation, a time characterized
by a small \( R_{*} \). Notice that the constant \( \hat{B} \) cancels
out in the perturbative differential equations, so its smallness plays
no role on the smallness of the extra terms in the perturbative differential
equations (\ref{eqvarphi}) and (\ref{eqchi}). These terms are small
because they involve derivatives of high degree and powers of the
scalar potential, which has a small coupling constant.

As for the behavior of the noncommutative tensor prior to \( t_{*} \),
we propose no model for \( B \)\textit{\emph{. Even if the expression
\( B=\hat{B}R^{-2} \) or other one with a singularity for \( B \)
at \( R=0 \) apply, this clearly occurs before perturbation theory
is valid. Furthermore, if \( t_{*} \) coincides with the onset of
inflation then the physics prior to \( t_{*} \) has small impact,
since chaotic initial conditions are assumed. }}

\textit{\emph{There is also another scenario where our conclusions
remain valid. If the nonperturbative regime allows for inflation then
it could be that part of inflation is initialy driven by noncommuativity
and, at a later time, driven by the mechanism here described. This
would allow a perturbative treatment beginning at a later time \( t_{*} \)
so that the scale factor \( R_{*} \) would be larger by several orders
of magnitude, and would allow larger values for \( \hat{B} \), also
by several orders of magnitude.}}

\section{\textit{\emph{Appendix}}}

\textit{\emph{Here we perform some of the explicit calculations leading
to the invariance of the theory under rotations, and illustrate the
small corrections that noncommutativity introduces in the slow-roll
regime. We recall that \( \partial _{i}\Phi =0 \) and \( \partial _{i}\theta ^{\alpha \beta }=0 \)
so that all spatial derivatives vanish. We use the notation \begin{equation}
\label{theta_param}
\theta ^{\alpha \beta }=\left( \begin{array}{cccc}
0 & -E_{x} & -E_{y} & -E_{z}\\
E_{x} & 0 & -B_{z} & B_{y}\\
E_{y} & B_{z} & 0 & -B_{x}\\
E_{z} & -B_{y} & B_{x} & 0
\end{array}\right) .
\end{equation}
 Our aim is to calculate, for instance \begin{equation}
\Omega \equiv \left[ \frac{1}{2}V''\theta ^{\alpha _{1}\beta _{1}}\theta ^{\alpha _{2}\beta _{2}}\Phi _{;\beta _{1}\beta _{2}}\right] _{;\alpha _{2};\alpha _{1}}=T^{\alpha _{1}\alpha _{2}}_{\qquad ;\alpha _{2}\alpha _{1}},
\end{equation}
 where the definition of the tensor \( T^{\alpha _{1}\alpha _{2}} \)
is evident.}}

\textit{\emph{It is easy to show that \begin{equation}
T^{\alpha _{1}\alpha _{2}}_{\qquad ;\alpha _{2}\alpha _{1}}={\partial _{\alpha _{1}}\left( \sqrt{-g}T_{\qquad ;\alpha _{2}}^{\alpha _{1}\alpha _{2}}\right) \over \sqrt{-g}}={\partial _{t}\left( R^{3}T_{\quad ;\alpha _{2}}^{t\alpha _{2}}\right) \over R^{3}},
\end{equation}
}}

\noindent \textit{\emph{and hence \begin{eqnarray}
T_{\quad ;\alpha _{2}}^{t\alpha _{2}} & = & {\partial _{\alpha _{2}}\left( \sqrt{-g}T^{t\alpha _{2}}\right) \over \sqrt{-g}}+\Gamma ^{t}_{\, \, \alpha _{2}\nu }T^{\alpha _{2}\nu }\nonumber \\
 & = & {\partial _{t}\left( R^{3}T^{tt}\right) \over R^{3}}+R\dot{R}\delta _{ij}T^{ij}.
\end{eqnarray}
}}

\textit{\emph{To evaluate the relevant entries of the tensor \( T^{\alpha _{1}\alpha _{2}} \)
we compute}}

\textit{\emph{\begin{equation}
S_{\mu \nu }\equiv \Phi _{;\nu \mu }=\ddot{\Phi }\delta _{\mu }^{t}\delta _{\nu }^{t}-R\dot{R}\dot{\Phi }\tilde{\delta }_{\mu \nu }
\end{equation}
 to get}}

\textit{\emph{\begin{eqnarray}
T^{tt} & = & -\frac{1}{2}R\dot{R}\dot{\Phi }V''\delta _{\mu \nu }\theta ^{t\mu }\theta ^{t\nu }=-\frac{1}{2}R\dot{R}\dot{\Phi }V''E^{2},\nonumber \\
\delta _{ij}T^{ij} & = & \frac{1}{2}V''\left[ \delta _{ij}\theta ^{it}\theta ^{it}\ddot{\Phi }-R\dot{R}\dot{\Phi }\delta _{ij}\theta ^{i\beta _{1}}\theta ^{j\beta _{2}}\delta _{\beta _{1}\beta _{2}}\right] \nonumber \\
 & = & \frac{1}{2}V''\left[ E^{2}\ddot{\Phi }-2R\dot{R}\dot{\Phi }B^{2}\right] .
\end{eqnarray}
}}

\textit{\emph{Thus we can explicitly verify the invariance of these
expressions under rotations as well as under translations.}}

\textit{\emph{Taking \( \vec{E}=0 \) leads to \begin{equation}
\Omega =-{1\over R^{3}}{\partial _{t}\left[ R^{5}\dot{R}^{2}\dot{\Phi }B^{2}V''\right] }.
\end{equation}
}}

\textit{\emph{Using perturbation theory in the slow-roll approximation,
Eqs. (\ref{slow_eqs}), (\ref{Fried}) and (\ref{slow_cond}), one
is lead to evaluate \begin{eqnarray}
\Omega ' & = & -{2\over a^{3}}{\partial _{t}\left[ a^{5-2\varepsilon }\dot{a}^{2}\dot{\phi }\frac{1}{2}V''\right] }\nonumber \\
 & = & -{16\pi \dot{\phi }\over 3a^{3}}{\partial _{t}\left[ a^{7-2\varepsilon }V\frac{1}{2}V''\right] }\\
 & = & -{16\pi a^{4-2\varepsilon }\dot{\phi }\over 3}{\left[ \left( 7-2\varepsilon \right) \sqrt{\frac{8\pi }{3}}V^{1/2}F+F'\dot{\phi }\right] },\nonumber 
\end{eqnarray}
 where \( F=\frac{1}{2}VV'' \). Using Eq. (\ref{maj_fipto}) and
\( \varepsilon =2 \), it follows that \begin{equation}
\left| \Omega '\right| \leq {16\sqrt{2}\pi \over 3}\left| V\right| \left[ 3\sqrt{\frac{8\pi }{3}}\left| F\right| +\sqrt{2}\left| F'\right| \right] .
\end{equation}
}}

\textit{\emph{Finally, from Eqs. (\ref{pot}) and (\ref{maj_pot}),
one obtains \begin{equation}
\left| \Omega '\right| \leq 2.4\times 10^{-37},
\end{equation}
 which illustrates how small are the corrections introduced by noncommutativity.}}

\end{document}